\documentclass[submission,copyright,creativecommons]{eptcs}
\providecommand{\event}{FMAS 2025} 

\usepackage{iftex}

\ifpdf
  \usepackage{underscore}         
  \usepackage[T1]{fontenc}        
\else
  \usepackage{breakurl}           
\fi
\usepackage{cite}
\usepackage{amsmath,amssymb,amsfonts}
\usepackage{algorithmic}
\usepackage{graphicx}
\usepackage{textcomp}
\usepackage{wrapfig}
\usepackage{xspace}
\usepackage[table,xcdraw]{xcolor} 
\usepackage{listings}
\usepackage{hyperref}
\usepackage[nameinlink,capitalise]{cleveref}

\newcommand{\wrap}[1]{\begin{tabular}{@{}c@{}}#1\end{tabular}}
\newcommand{\Lince}{\textsf{Lince}\xspace}

\title{Analyzing Many Simulations of Hybrid Programs in \Lince}
\author{Reydel Arrieta
\ttfamily\institute{CISTER, Polytechnic Institute of Porto \\ Portugal}
\ttfamily\email{arrie@isep.ipp.pt}
\and
José Proença
\ttfamily\institute{CISTER \& University of Porto\\
Portugal}
\ttfamily\email{\quad jose.proenca@fc.up.pt}
\and
Patrick Meumeu Yomsi
\ttfamily\institute{CISTER, Polytechnic Institute of Porto \\
Portugal}
\ttfamily\email{pmy@isep.ipp.pt} 
}
\def\titlerunning{Analyzing Many Simulations of Hybrid Programs in \Lince}
\def\authorrunning{Reydel Arrieta, José Proença \& Patrick Meumeu Yomsi}
\begin{document}
\maketitle

\begin{abstract}
\textbf{Abstract.} Hybrid systems are increasingly used in critical applications such as medical devices, infrastructure systems, and autonomous vehicles. 
\Lince is an academic tool for specifying and simulating such systems using a C-like language with differential equations. 
This paper presents recent experiments that enhance \Lince with mechanisms for executing multiple simulation variants and generating histograms that quantify the frequency with which a given property holds.
We illustrate our extended \Lince using variations of an adaptive cruise control system.
\end{abstract}

\section{Contextualization and Motivation} \label{sec:motivation}
\vspace{-7pt}
Hybrid systems are complex systems that combine discrete with continuous behavior, where it is essential to meet timing, safety, functional and non-functional requirements under different operating conditions~\cite{laplante2022requirements}. These systems are commonplace in applications such as cyber-physical systems, robotics, automotive and industrial automation.   Several well-known tools like Uppaal~\cite{david2015uppaal} and KeYmaera~\cite{platzer2008keymaera} have been proposed in the literature to analyze their behavior and ensure their correctness. Although powerful, these tools struggle with non-linear systems and computational complexity.
\Lince~\cite{mendes2024formal,DBLP:conf/ppdp/0001N19,neves2018hybrid}
is a newer (academic) tool to simulate hybrid systems using a simple web-based interface, targeting programs with variables that can evolve based on systems of differential equations.

Currently, \Lince relies primarily on trajectory plots for analysis. However, assessing the impact of changes in configuration parameters or scenarios requires manually generating and checking each variation individually. This work investigates how to analyze multiple simulations, either by overlapping many variations in a single plot, or by counting how many times a given property holds over time.

\begin{figure}
    \centering
    \includegraphics[scale=0.5]{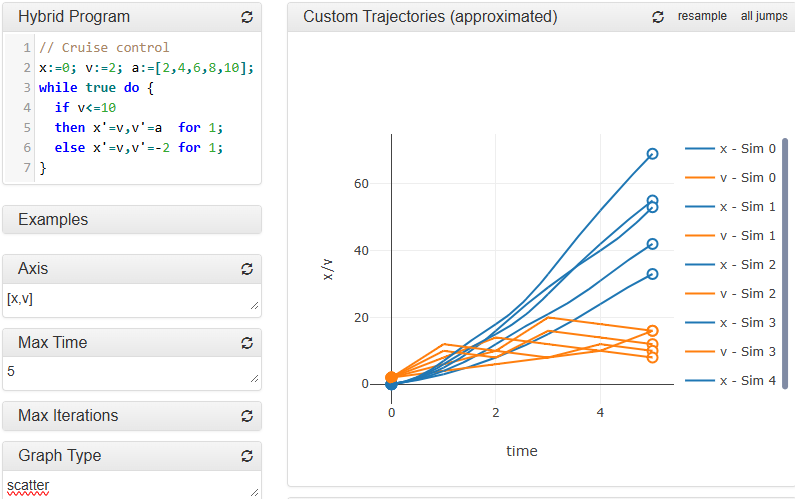}
    \caption{Modelling of a cruise control system in \Lince}
    \label{fig:exampleACC}
\end{figure}

\smallskip
\noindent
\textbf{Motivating example.}
Consider a simple cruise control system.
The interface of \Lince for this example is depicted in \cref{fig:exampleACC}:
the program is specified in the top-left box, other parameters are
configured in the bottom-left boxes, and the resulting plot is produced on the right. This hybrid program models the temporal evolution of a car's position \textbf{x} and velocity \textbf{v} by solving a system of first-order ordinary differential equations (ODEs).

The system is initialized with position $\mathbf{x:=0}$ and velocity $\mathbf{v:=2}$; while acceleration values are selected from the array $\mathbf{a:=[2,4,6,8,10]}$. A separate simulation is carried out for each acceleration value. In each case, the car evolves with constant acceleration~$\mathbf{a}$, while its velocity fulfils $\mathbf{v \leq 10}$.
Every time unit, the system checks if the velocity exceeds the threshold $\mathbf{10}$. If so, the system brakes with a deceleration of $\mathbf{-2}$.

Overlapping trajectories can make it difficult to verify velocity constraints or detect abnormal accelerations, e.g., \emph{ensuring that the vehicle never exceeds a given velocity}. To overcome this limitation, we propose an approach for analyzing multiple simulations simultaneously. This method evaluates whether a given property holds across different time intervals and aggregates the results, which are then visualized using histogram charts to provide an intuitive representation of how frequently the property is satisfied.

\smallskip\noindent
\textbf{Contributions.}
The core contribution of this paper is twofold:
($1$)~a step-by-step guide to using \Lince through a use case involving a  variant of an adaptive cruise control~(ACC) system\,footnote{ACC systems aim to maintain a safe distance between vehicles by dynamically adapting the acceleration of a following vehicle to the behavior of the vehicle in front.} and
($2$)~an extension of \Lince that facilitates the analysis of multiple simulations of a single system with different configurations. 
The goal of this extension is not to provide a set of reusable libraries for multiple simulations and visualisation of complex systems, such as Numpy libraries for Python,\footnote{https://numpy.org/} but to enrich Lince with the capability of running, analysing, and visualising many simulations.
In particular, we extend \Lince by supporting histograms~\cite{gedcke2001histogramming} that count the number of times a given goal is reached at different points in time, and enrich the language to describe ranges of possible configurations.
Using this extension, \Lince can now run a system multiple times and produce
a graph that marks, for each point in time from a set of sampling points,
how often a given (desirable or undesirable) state is reached.

\smallskip\noindent
\textbf{Paper structure.}
The paper is structured as follows. \Cref{sec:design} details the design and implementation of \Lince, covering the tool's architecture and the new histogram functionality for temporal analysis. \Cref{sec:cc} shows
the application of \Lince using an ACC case study, highlighting standard simulation and histogram-based analysis. \Cref{sec:conclusion} concludes the paper and describes future work.
\break

\vspace{-5pt}
\section{Design and Implementation} \label{sec:design}
\vspace{-5pt}
This section describes the internal architecture of \Lince and highlights recent extensions that aim to improve its analytical capabilities. 
It then describes our extension with
histogram-based analysis, providing more intuitive insights into temporal patterns of system behavior.

\subsection{Overview of Lince}
Lince is implemented in Scala, compiles to JavaScript, features a web-based front-end, and includes a server for symbolic computations (though this server is not exploited in this work, which uses the SageMath algebraic solver) to avoid approximations.
Lince is a lightweight and easily extensible tool, but not yet as mature or scalable as existing alternatives such as Simulink\cite{sanfelice2013toolbox} or Uppaal\cite{DBLP:journals/corr/abs-1207-1272,david2015uppaal}. Recent enhancements improved usability, methods, and visualization~\cite{mendes2024formal}, making it more practical. These improvements have enabled researchers to model more complex interactions between computational and physical components effectively, thus advancing the study and application of hybrid systems.

\subsection{Lince Extension for Histogram Generation}
\begin{figure}[h]
    \centering
    \wrap{\includegraphics[width=0.5\textwidth]{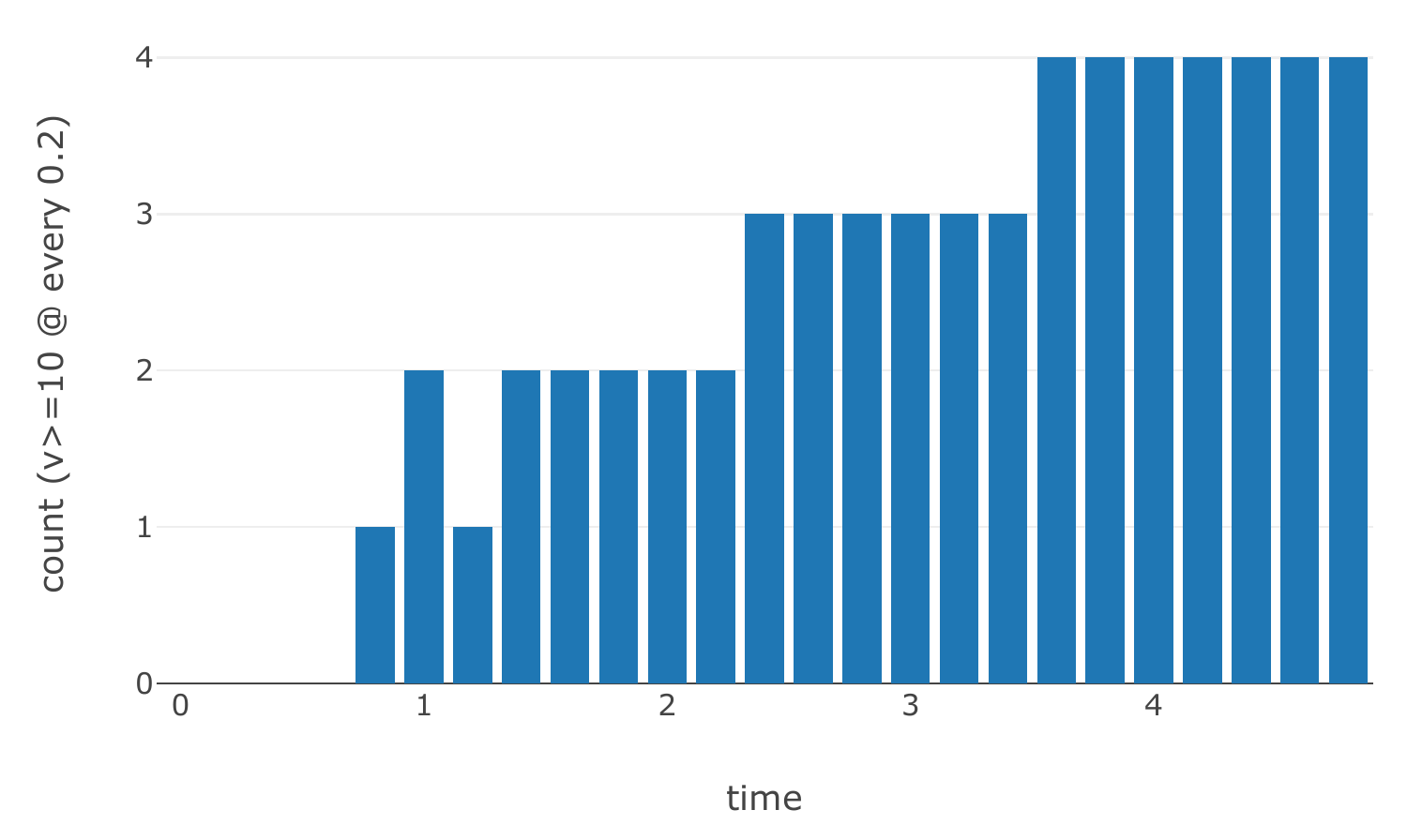}}~
    \wrap{\includegraphics[width=0.5\textwidth]{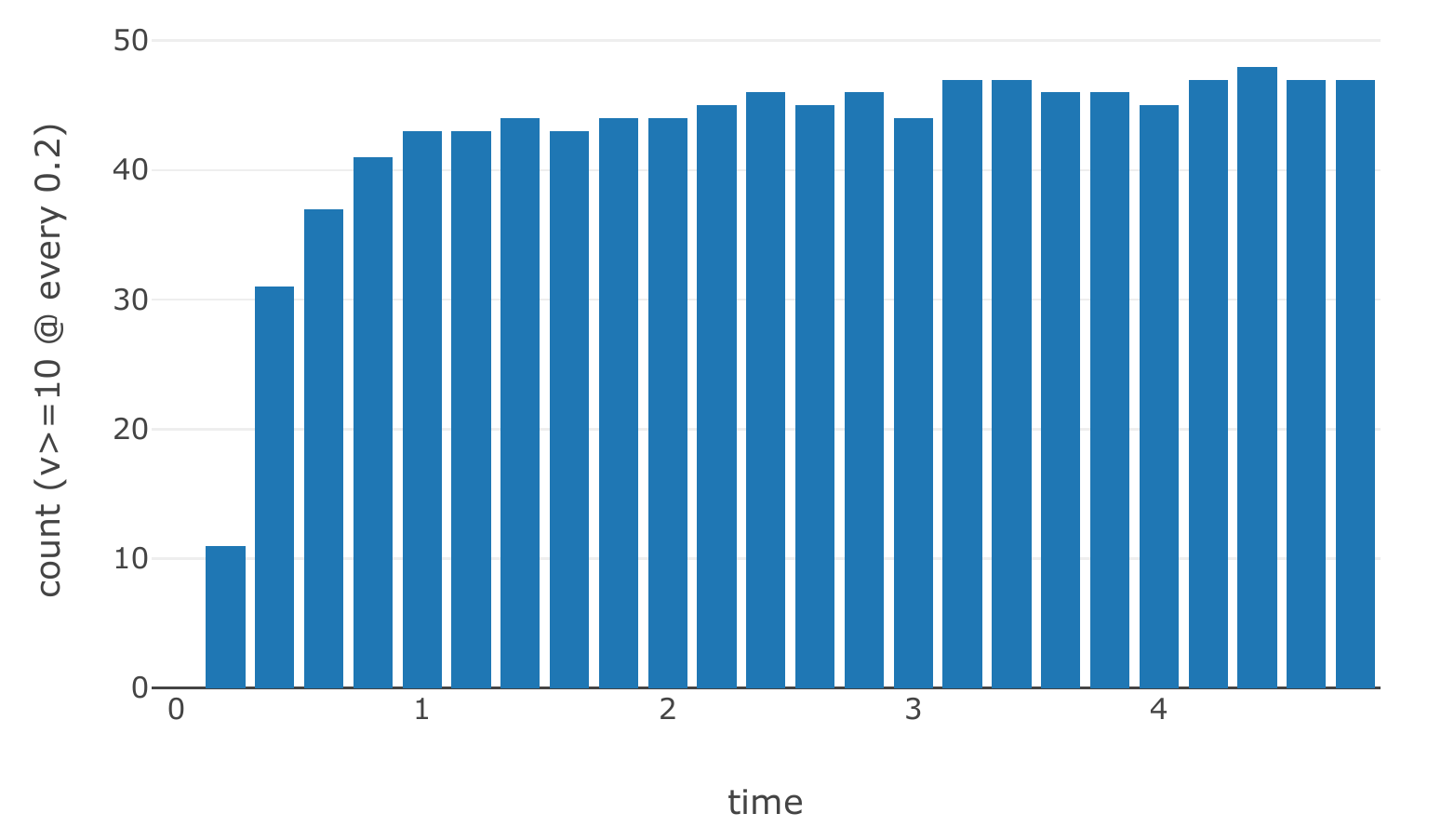}}
   \caption{ Histogram plotting of Cruise Control with 5 and 50 simulations, respectively}
    \label{fig:CCexample}
\end{figure}

\vspace{-5pt}
Histograms depict how often a particular goal is achieved over time,
allowing users to observe temporal patterns in system behavior.
This functionality in Lince was inspired by similar features in tools such as Uppaal~\cite{david2015uppaal}.
In \Lince, it has been adapted to support the specific needs of hybrid systems analysis, providing a clear visualization of requirements' compliance over time.
To generate a histogram in Lince, users modify the \textit{Graph type} field by replacing the \textbf{scatter} command with \textbf{histogram} command (c.f. \cref{fig:exampleACC}. The format of a histogram command is \texttt{histogram: <requirement> @ <sampling>}, where \texttt{requirement} describes the condition of interest to be checked, e.g., \texttt{v>=10}, and \texttt{sampling} is an optional argument describing the points used to evaluate the requirement, e.g., the sampling \texttt{step 5} divides the run time into 5 evenly distributed sampling points, and \texttt{every 0.2} sets sampling points every 0.2 time units). \cref{fig:CCexample} illustrates two histograms obtained from the example in \cref{fig:exampleACC} using 5 and 50 different acceleration values, respectively.

The implementation of this functionality was carried out in two main phases: ($1$)~the adaptation of Lince to collect relevant data during the analysis, and ($2$)~the generation and graphical visualization of histograms from the collected data. This enhancement not only improves the user experience, but also extends Lince’s analytical capabilities, enabling a more in-depth study of temporal patterns in the fulfillment of requirements.

\section{Analyzing an Adaptive Cruise Control} \label{sec:cc}

\newcommand{\ld}[1]{%
    \ensuremath{\textcolor{orange!75!black}{\mathsf{#1_{\ell}}}}\xspace}
\newcommand{\fl}[1]{%
    \ensuremath{\textcolor{green!65!black}{\mathsf{#1_{f}}}}\xspace}
\newcommand{\ldr}{\textcolor{orange!75!black}{\ell}}
\newcommand{\flw}{\textcolor{green!65!black}{\mathsf{f}}}

 This section presents a formal model of an ACC system that uses a variant of a hybrid program in \Lince,\footnote{ The source code of Lince is available online at \href{https://github.com/arcalab/lince}{https://github.com/arcalab/lince.}} and describes how a safe distance to the vehicle ahead is maintained (and a target speed is ignored).
This use case can be run online at \textbf{http://arcatools.org/lince}.

\medskip

\begin{wrapfigure}{r}{0.4\textwidth}
\begin{lstlisting}[language=Python,morekeywords={then,do}]
while true do {
  if Safe(pf,vf,pl,vl,al)
  then af:=bwd;
  else af:=fwd;
pf'=vf,vf'=af,af'=0,
pl'=vl,vl'=al,al'=0 for st;}
\end{lstlisting}
\end{wrapfigure}

\noindent
\textbf{ACC modeling.}
We consider a \Lince hybrid program that simulates two vehicles on a straight road: a \textit{$\ldr$eader} and a \textit{$\flw$ollower}.
The core of the ACC~program is the predicate \textsf{Safe}(\fl{p}, \fl{v}, \ld{p}, \ld{v}, \ld{a}), which determines whether it is safe for the \textit{follower} to accelerate. In particular, it checks whether, given the position \fl{p} and velocity \fl{v}, of the \textit{follower} together with the position \ld{p}, velocity \ld{v}, and acceleration \ld{a} of the \textit{leader}, a collision would occur if the \textit{follower} accelerates during the next sample time~(\textbf{st}).
The program includes an infinite while-loop where, at each iteration, the dynamics of both vehicles evolve according to their differential equations. Inside the loop, the \textsf{Safe} function is invoked to guide decisions about acceleration or braking. 

To compute the \textsf{Safe} predicate, we assume that the \textit{follower} starts by accelerating for a period of $\mathbf{st}$ (\emph{first phase}), and then brakes (\emph{second phase}). We then manually compute whether the vehicles collide in the \emph{first} or \emph{second} phase. If any collision is detected, \textbf{Safe} returns false and the \textit{follower} sets its acceleration to $\mathbf{bwd} \in \mathbb{R}^{-}$ (since it is not safe to accelerate); otherwise it returns true and the \textit{follower} sets its acceleration to $\mathbf{fwd} \in \mathbb{R}^{+}_{*}$. Hereafter, we explain the dynamics of each phase.

\newcommand{\deltat}{\textbf{st}\xspace}

\medskip
\noindent\textbf{First phase.}
Assuming the \textit{leader} maintains a constant acceleration \ld{a}, the estimated positions of the \textit{follower} $\fl{p^{\mathbf{st}}}$ and the \textit{leader} $\ld{p^{\mathbf{st}}}$ up to time \textbf{st} are given by $\fl{p^{\mathbf{st}}} = \tfrac{\mathbf{fwd}}{2}\cdot \deltat^2 + \fl{v}\cdot \deltat + \fl{p}$ and $\ld{p^{\mathbf{st}}} = \tfrac{\ld{a}}{2}\cdot \deltat^2 + \ld{v}\cdot \deltat + \ld{p}$, respectively. 
Furthermore, the \textit{follower}'s velocity is $\fl{v^{\mathbf{st}}} = \mathbf{fwd}\cdot \deltat+\fl{v}$.
Therefore, the system reduces to a numeric identity, since there are no variables. As a result, the positions of the vehicles after a time step $\mathbf{st}$ can be determined exactly.
%
If $\fl{p^{\mathbf{st}}} < \ld{p^{\mathbf{st}}}$, then we assume that the vehicles did not collide in the first phase. 

\medskip
\noindent\textbf{Second phase.} 
After \textbf{st} time units, we check if the trajectory $\fl{p}(t)$ of the \textit{follower} can intersect the trajectory $\ld{p}(t)$ of the \textit{leader} at any time $t\geq 0$. 
More specifically, we estimate these trajectories as follows, where \textbf{fwd} and \textbf{bwd} are the acceleration and braking values of the \textit{follower}, respectively, and $\fl{p^{\mathbf{st}}}$, $\ld{p^{\mathbf{st}}}$ and \fl{v^{\mathbf{st}}} are defined above.
$$
\fl{p}(t) = \frac{\mathbf{bwd}}{2}\cdot t^2 + \fl{v^{\mathbf{st}}} \cdot t
       +  
          \fl{p^{\mathbf{st}}}
    \;\; \text{and}
    \;\;
    \ld{p}(t) = 
                 \frac{\ld{a}}{2}\cdot t^2+\ld{v}\cdot t + \ld{p^{\mathbf{st}}}
$$

\noindent
A collision occurs if there exists a time $t$ such that $\ld{p}(t) - \fl{p}(t) = 0$, which evaluates to the following equation.

\begin{multline*}
    \underbrace{\frac{(\ld{a}-\mathbf{bwd})}{2}\cdot}_{at}  t^2+\underbrace{\left[(\ld{a}-\mathbf{fwd})\cdot(\deltat)+(\ld{v}-\fl{v})\right]}_{bt}\cdot t+ \\
    \underbrace{\left[\frac{(\ld{a}-\mathbf{fwd})}{2}\cdot(\deltat)^2+(\ld{v}-\fl{v})\cdot(\deltat)+ (\ld{p}-\fl{p})\right]}_{ct}=0
\end{multline*}
We then search for solutions, which exist if $\Delta = bt^2 - 4\cdot at\cdot ct \geq 0$ or $at==0$ and $t=-ct/bt>0$.
    
\medskip
\noindent\textbf{Combining both phases.}

Combining the results above, the final definition of \textsf{Safe} is defined as follows.

\begin{multline*}
    \textsf{Safe}(\fl{p}, \fl{v}, \ld{p}, \ld{v}, \ld{a})= 
    \ld{p}(\deltat)\le\fl{p}(\deltat) \lor \left [\ld{a}==\mathbf{bwd} \land -ct/bt>0\right] \lor \\ \left[\Delta\ge0 \land \ld{a}\neq \mathbf{bwd} \land (\frac{-bt + \sqrt{\Delta}}{2\cdot at} > 0 \lor \frac{-bt - \sqrt{\Delta}}{2\cdot at} > 0)\right] 
\end{multline*} 

\medskip\noindent
\textbf{Simulation setups and analysis of the results.}
In our simulations, we use $\mathbf{fwd}=3$, $\mathbf{bwd}=-3$, and $\mathbf{st}=2$.
Note that $\mathbf{st}=2$ reflects a value commonly used in the automotive industry~\cite{rajamani2011vehicle,piao2008advanced}, where the prediction horizons for forward position and collision prediction are typically between 2 and 4 seconds.
The \textit{leader} is assumed to be stationary ($\ld{v}=0$) at $\ld{p} = 50$ units from the origin. A set of integer values for acceleration \ld{a} is defined, where each value corresponds to a separate run. This set is denoted 
as $\ld{a}:= [-3..3]$, which represents a discrete sampling of possible acceleration profiles and constitutes a basic contribution to the development of \Lince. The \textit{follower}, in contrast, is positioned at the origin, also at rest, and begins with a constant acceleration $\fl{a}=3$. The system dynamics are modeled by an infinite loop (\texttt{while true}), in which the states of the two vehicles evolve according to differential equations that are integrated over the sampling interval \textbf{st:=2}. Specifically, for the \textit{follower}: \texttt{\fl{p}'=\fl{v}, \fl{v}'=\fl{a}} and for the \textit{leader}: \texttt{\ld{p}'=\ld{v}, \ld{v}'=\ld{a}}.

\begin{figure}[!h]
    \centering
    \wrap{\includegraphics[width=0.4\textwidth]{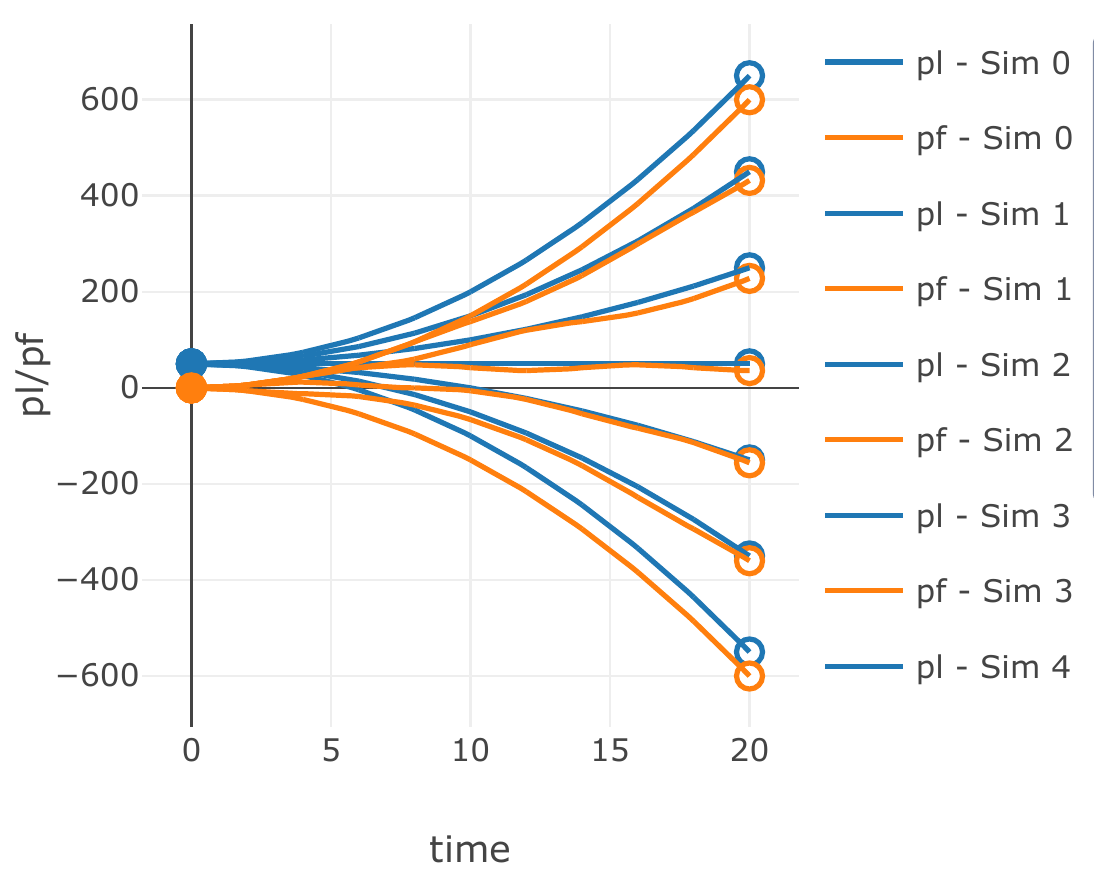}}~
    \wrap{\includegraphics[width=0.5\textwidth]{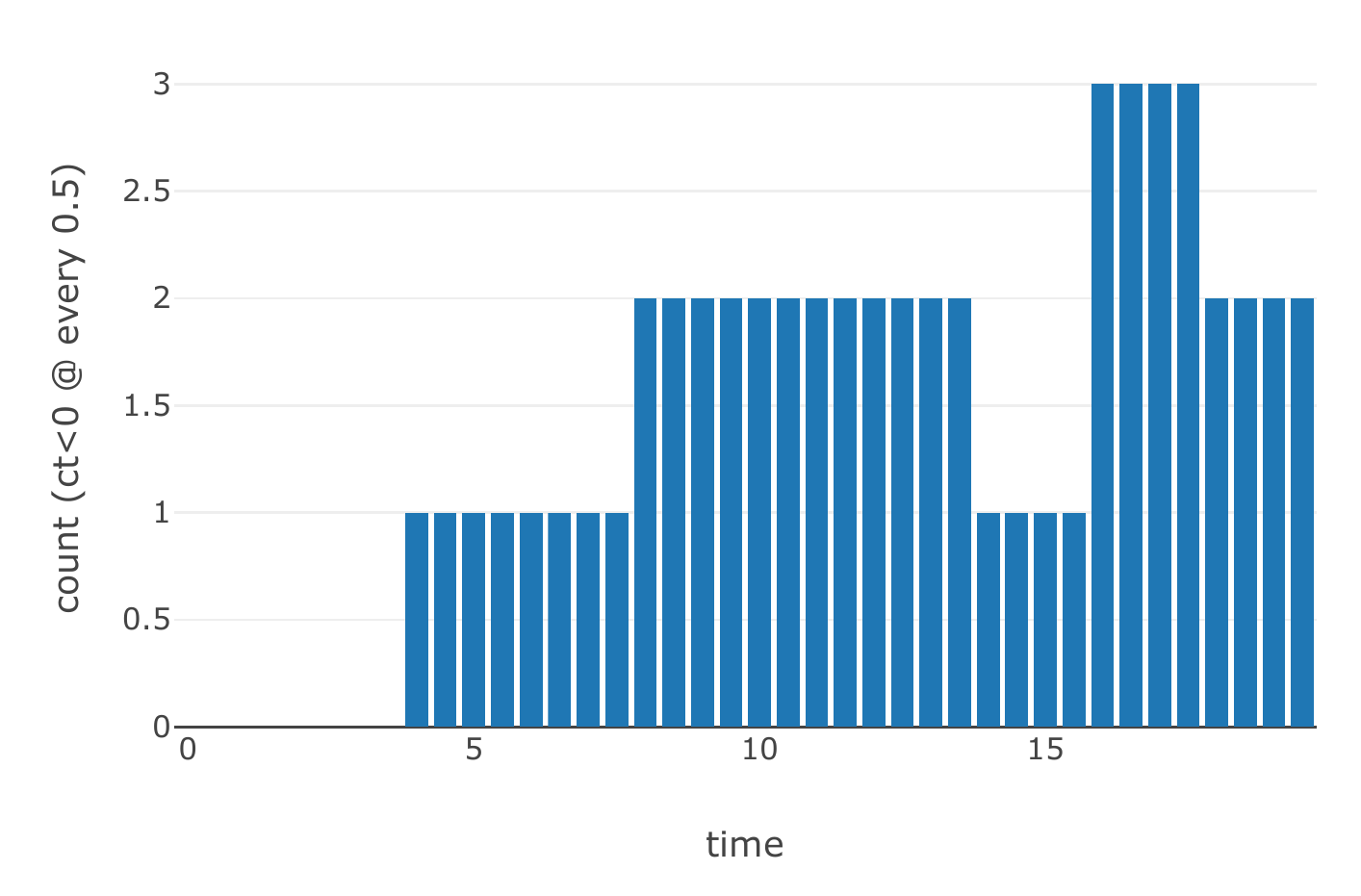}}
   \caption{Simulations of the ACC in \Lince, displaying the trajectories of the vehicles (left) and the histogram counting when $ct\le0$.}.
    \label{fig:ACCexample}
\end{figure}

After modeling the scenario in \Lince, a complex graph is obtained consisting of seven simulation traces, each corresponding to one of the previously defined acceleration values (see Figure~\ref{fig:ACCexample}, left). 
Overlapping trajectories make braking events difficult to identify.
To address this, our approach introduces the use of the \textit{Graph type} field with the option \textbf{ histogram: $ct\le0$ @ every 0.5}, for example. Here we determine how many collisions occur during the first phase in particular, i.e., $\ld{p}(\deltat)\le\fl{p}(\deltat)$.  This way yields a more interpretable visualization (see Figure~\ref{fig:ACCexample}, right). 
It is important to note that this constitutes a discretized analysis; therefore, reducing the evaluation interval increases the accuracy of the results but also introduces a higher computational complexity.  
The histogram is interpreted as follows: each increment represents the occurrence of a new event, while each decrement indicates the absence of an event. Periods without changes indicate steady states that correspond to previously established conditions and therefore do not convey additional information.

\noindent\textbf{The ACC use case analyzed by other tools.}
In \textbf{Simulink}~\cite{sanfelice2013toolbox}, our AAC use case could be modeled by creating multiple interconnected blocks for sensors, controllers, and vehicle dynamics, with parameters manually configured for each simulation run. This approach requires some familiarisation with these tools and how to properly configure all parameters. In \textbf{UPPAAL}~\cite{david2015uppaal}, the system would be represented as a network of timed automata, where each component (\textit{leader}, \textit{follower}, and controller) would be described through discrete states and clock constraints. Although UPPAAL targets timed automata (without ODEs)~\cite{behrmann2004uppaal}, it also supports simple ODEs such as the ones used in this ACC case, but not more complex ODEs (unlike Lince). UPPAAL further supports running multiple random simulations, but there is no explicit mechanism to build specific scenarios, and it supports a variety of queries to infer statistical analysis or produce plots and histograms.
Unlike Simulink and UPPAAL, Lince aims at being an easy to run tool, which does not require any installation (a web-browser is enough), and easy to extend and experiment, with a compact dedicated input language that is precise and rich enough to describe and analyse non-trivial examples. Lince does not aim at being efficient, scalable, nor providing an extensive set of features.
Table~\ref{tab:comparisontools} summarizes the qualitative differences between Simulink, UPPAAL, and Lince.

Using dedicated libraries, such as \textbf{Numpy}
for Python, one can produce most of these analysis by manually describing the system of equations and encoding all the control logics into a Python script. We believe that this is harder to write for non-experts than a Lince program, and harder to maintain and experiment, given that the declarative hybrid program (read by Lince) would be manually implemented in a Python script. Numpy (or an alternative) could be used as a possible back-end for Lince, although we currently focus on Scala libraries or JavaScript (JS) libraries (e.g., we use the Plotly JS libraries\footnote{\url{https://plotly.com/javascript/}} to draw plots and histograms), given that Lince is implemented in Scala and compiled to JS.

\newcommand{\wrapl}[1]{\begin{tabular}{@{}l@{}}#1\end{tabular}}
\begin{table}[!h]
\centering
\begin{tabular}{|l|p{4.0cm}|p{3.5cm}|p{3.8cm}|}
\hline
\textbf{Feature} & \textbf{Simulink} & \textbf{UPPAAL} & \textbf{Lince} \\ \hline

\wrapl{Model\\representation} & 
Graphical block diagrams & 
Networks of timed automata & 
Textual hybrid programs (guarded commands + ODEs) \\ \hline

Installation & 
Requires MATLAB environment & 
Requires desktop installation and modeling tool & 
Runs directly in browser \\ \hline

Focus & 
Large-scale numeric simulation and control design & 
Rigorous verification of real-time systems & 
Lightweight exploration of hybrid system logic \\ \hline

\wrapl{Model\\representation} & 
Not supported natively & 
Symbolic reasoning over guards and invariants of timed automata & 
Symbolic reasoning over guards, invariants, and continuous dynamics (ODEs) \\ \hline

User interaction & 
Graphical modeling through block connections; parameter tuning via GUI & 
Graphical editor for automata construction and query interface for verification & 
Direct code editing and instant visualization of trajectories and histograms in the browser \\ \hline
\end{tabular}

\caption{Comparison of modeling and analysis features across tools.}
\label{tab:comparisontools}
\end{table}

\break

\medskip
\noindent\textbf{Lessons learned.}
Guided by our ACC example, we identified difficulties in the modelling and analysis of non-trivial CPS, namely regarding the impact of experimenting with variations of some parameters. In this example, it was not clear how often a vehicle would reach the target proximity of a leading vehicle in different scenarios with different leading vehicles. Other experiments not reported here, involving injected errors in the periodicity delay, further supported the usefulness of quickly experimenting and simulating these kinds of critical scenarios.
We also concluded that further analysis or extensions could be useful, such as: calculating expected values, supporting modular definitions of each vehicle in separate, or supporting fine-tune configuration of some hard-coded parameters (e.g., the precision of the ODE solver). This is left for future work.

\section{Conclusion and Future Work} \label{sec:conclusion}
\vspace{-5pt}
Analyzing hybrid systems poses significant challenges due to the combination of discrete and continuous behaviors, especially in critical applications where temporal requirements are essential. To address this issue, \Lince has evolved by incorporating new functionalities that facilitate the interpretation of massive simulations and the verification of temporal properties. The main contribution of this improvement is the integration of histograms, which provide a clear visualization of the frequency with which certain events occur over time. This resolves the previous problem of graph overlapping in multiple simulations, offering a direct way to detect compliance patterns and anomalies. Furthermore, adopting a simplified notation for defining lists with unit-step discretization has optimized the simulation setup process, making the tool more agile.

These results open an interesting avenue for future work: extending the current support for random value generation towards a principled treatment of stochastic hybrid systems. Our recent paper~\cite{PPDP25-lince} introduces a while language that integrates probabilistic constructs with differential dynamics, and shows how Lince can be extended to reason about stochastic hybrid programs. Experimental histograms can already be found in that paper, which are formally presented here.

We are also exploring how Lince could be used to validate formal monitors or to generate tests and monitors from logical specifications, e.g., inspired on the work by Yamaguchi et al.~\cite{YamaguchiHN24}.
On a different front, we plan to investigate modularity with existing toolchains and mathematical modelling tools, and modularity of the specifications, supporting the composition of independent components.

\section{Acknowledgments}\label{sec:ack}
This work was supported by national funds through FCT/MECI (Portuguese Foundation for Science and Technology), within the CISTER Research Unit (UIDP/UIDB/04234/2020) and under the project Intelligent Systems Associate Laboratory - LASI (LA/P/0104/2020) and Project Route 25 (ref. TRB/2022/00061 - C645463824-00000063), funded by the EU/Next Generation, within call n.º 02/C05-i01/2022 of the Recovery and Resilience Plan (RRP)

\break

\bibliographystyle{eptcs}
\bibliography{bib/david, bib/jose,bib/other}

\end{document}